\documentclass[aps,prd,reprint,superscriptaddress,showpacs,amsmath,amssymb]{revtex4-2}

\usepackage[colorlinks,
           bookmarks=true,
           linkcolor=blue,
           urlcolor=blue, 
            anchorcolor=black,
            citecolor=blue
            ]{hyperref}

\usepackage{multirow}
\usepackage{graphicx}
\usepackage{dcolumn}
\usepackage{bm}
\usepackage{float}
\usepackage{exscale}
\usepackage{relsize}
\usepackage{amsmath}
\usepackage{amssymb}
\usepackage{placeins}
\begin{document}

\title{Shedding light on the nature of $\phi(2170)$ with the parton and hadron cascade model PACIAE}
\author{Jian Cao}
\affiliation{School of Physics and Information Technology, Shaanxi Normal University, Xi'an 710119, China}
\author{Wen-Chao Zhang}
\email{wenchao.zhang@snnu.edu.cn (corresponding author)}
\affiliation{School of Physics and Information Technology, Shaanxi Normal University, Xi'an 710119, China}
\author{Bo Feng}
\affiliation{School of Physics and Information Technology, Shaanxi Normal University, Xi'an 710119, China}
\author{Ya-Hui Hou}
\affiliation{School of Physics and Information Technology, Shaanxi Normal University, Xi'an 710119, China}
\author{An-Ke Lei}
\affiliation{School of Physics and Electronic Science, Guizhou Normal University, Guiyang, 550025, China}
\author{Zhi-Lei She}     
\affiliation{Wuhan Textile University, Wuhan 430200, China}
\author{Hua Zheng}
\affiliation{School of Physics and Information Technology, Shaanxi Normal University, Xi'an 710119, China}
\author{Li-Lin Zhu}
\affiliation{College of Physics, Sichuan University, Chengdu 610064, China}
\author{Dai-Mei Zhou}
\email{zhoudm@mail.ccnu.edu.cn}
\affiliation{Key Laboratory of Quark and Lepton Physics (MOE) and Institute of
            Particle Physics, Central China Normal University, Wuhan 430079,
            China}
\author{Yu-Liang Yan}
\affiliation{China Institute of Atomic Energy, P. O. Box 275 (10), Beijing
            102413, China}
\author{Ben-Hao Sa}
\email{sabhliuym35@qq.com} 
\affiliation{China Institute of Atomic Energy, P. O. Box 275 (10), Beijing
            102413, China}   
\affiliation{Key Laboratory of Quark and Lepton Physics (MOE) and Institute of
            Particle Physics, Central China Normal University, Wuhan 430079,
            China}
\date{\today}

\begin{abstract}

The nature of $\phi(2170)$  remains an open question in hadron spectroscopy. In this work, we simulate its production in $e^+e^-$ collisions at $\sqrt{s}=4.95$ GeV using the parton and hadron cascade model PACIAE 4.0, which sequentially generates the final partonic state (FPS) and the final hadronic state (FHS). While previous studies have interpreted $\phi(2170)$ as an $ss\bar{s}\bar{s}$ or a $u\bar{u}s\bar{s}$ tetraquark state, the $U(1)$ anomaly coupling allows non-strange quarks to couple to a vector $s\bar{s}$ component via soft-gluon interactions. This motivates us to also explore the $d\bar{d}s\bar{s}$ tetraquark configuration. In addition, we consider $\phi(2170)$ as an excited strangeonium state, an $s\bar{s}g$ hybrid state, a $\bar{\Lambda}\Lambda$ bound state, and a $\phi K^+K^-$ resonance state. The strangeonium, hybrid, and tetraquark candidates are formed by coalescing their constituent partons in the FPS using the dynamically constrained phase-space coalescence model. The $\bar{\Lambda}\Lambda$ and $\phi K^+K^-$ states are produced via recombination of their constituent hadrons in the FHS. We calculate the orbital angular momentum quantum number of each candidate in its rest frame and perform spectral classification. Given $J^{PC}=1^{--}$,  $\phi(2170)$ can be interpreted as a $D$-wave $s\bar{s}$, a $P$-wave $s\bar{s}g$, a $P$-wave $u\bar{u}s\bar{s}/d\bar{d}s\bar{s}/ss\bar{s}\bar{s}$, an $S$-wave $\bar{\Lambda}\Lambda$, or an $S$-wave $\phi K^+K^-$ state. We estimate the production yields for all these configurations: those of the $D$-wave $s\bar{s}$, $P$-wave $s\bar{s}g$, $u\bar{u}s\bar{s}$, and $d\bar{d}s\bar{s}$ states are of order $10^{-4}$; those for the $S$-wave $\bar{\Lambda}\Lambda$ and $\phi K^+K^-$ states are of order $10^{-5}$; while the $P$-wave $ss\bar{s}\bar{s}$ yield is of order $10^{-6}$. Moreover, significant discrepancies are observed in the rapidity distributions and the transverse momentum spectra among the various candidates. These discrepancies could serve as valuable criteria for unraveling the nature of $\phi(2170)$.

\end{abstract}

\maketitle

\section{Introduction} 

A central challenge in modern physics is understanding how quarks and gluons bind into hadrons. The vast majority of observed hadrons are either quark-antiquark pairs (mesons) or three-quark systems (baryons). The non-Abelian property of quantum chromodynamics (QCD) permits the existence of new types of hadrons, such as glueballs, hybrids, and multiquark states \cite{exotic_1, exotic_2, exotic_3}. These exotic hadrons provide a distinctive environment to investigate the strong interactions and the confinement mechanism \cite{exotic_meson}. The first exotic hadron, $X(3872)$, was discovered by the Belle Collaboration in $e^{+}e^{-}$ collisions in 2003. Since then, several exotic candidates consistent with tetraquark interpretations have been observed, including $X(2900)$ \cite{X2900}, $T_{cc}^+(3875)$ \cite{Tcc3900}, $Z_c(3900)$ \cite{Zc3900}, $X(6900)$ \cite{X6900_1, X6900_2, X6900_3}. Among these, $X(6900)$ is interpreted as a fully charm tetraquark state $cc\bar{c}\bar{c}$, with a mass around 6.9 GeV/c$^2$ and an enhanced signal in the $J/\psi J/\psi$ decay channel \cite{X6900_tetra_1, X6900_tetra_2, X6900_tetra_3, X6900_tetra_4, X6900_tetra_5}. Based on flavour symmetry and heavy‑quark symmetry in QCD, it is natural to conjecture that analogous states, a fully bottom tetraquark $bb\bar{b}\bar{b}$ and a fully strange tetraquark $ss\bar{s}\bar{s}$ should also exist.

Recently, the BESIII Collaboration reported the observation of an axial-vector particle, $X(2300)$, in the $\phi\eta'$ and $\phi\eta$ invariant mass spectra from the decay $\psi(3686)\to \phi\eta\eta'$ \cite{x_2300_BESIII}. Its mass and width are measured to be $2316\,\text{MeV}/c^2$ and $89\,\text{MeV}$, respectively, with spin-parity $J^{PC}=1^{+-}$ \cite{x_2300_BESIII}. In the same decay process, another state, $\phi(2170)$ (also known as $Y(2175)$), was observed with a mass of $2164\,\text{MeV}/c^2$, a width of $106\,\text{MeV}$, and quantum numbers $J^{PC}=1^{--}$ \cite{phi_2170}.
Although the parity of $\phi(2170)$ is different from that of $X(2300)$, the production rate of $\phi(2170)$ in the $\phi\eta'$ channel is found to be comparable to that of $X(2300)$. This comparable production rate in the same decay process suggests either a similar coupling mechanism to $\psi(3686)$ or an accidental coincidence of the partial widths to the $\phi\eta^{\prime}$ final states, which requires further investigation. $\phi(2170)$ was first observed by the BaBar Collaboration in 2006 via the initial-state radiation (ISR) process $e^+e^- \to \gamma_{\text{ISR}} \phi f_0(980)$ in the $\phi f_0(980)$ channel \cite{phi_2170_1}. Its observation was subsequently confirmed by the Belle Collaboration \cite{phi_2170_2}, the BES and BESIII Collaborations \cite{phi_2170,phi_2170_3,phi_2170_4,phi_2170_5,phi_2170_6,phi_2170_7}, and further supported by later BaBar analyses \cite{phi_2170_8, phi_2170_9}. Various theoretical interpretations have been proposed for the nature of $\phi(2170)$. These include an excited $s\bar{s}$ state with the $2^3D_1$ or $3^3S_1$ configuration \cite{phi_2170_exp1, phi_2170_exp1_1,phi_2170_exp1_2,phi_2170_exp2,phi_2170_exp3_1}, an $s\bar{s}g$ hybrid state \cite{phi_2170_exp3,phi_2170_exp4, phi_2170_exp4_1}, an $ss\bar{s}\bar{s}$ \cite{phi_2170_exp5,phi_2170_exp6,phi_2170_exp7,phi_2170_exp8,phi_2170_exp9} or a $u\bar{u}s\bar{s}$ \cite{phi_2170_exp10} tetraquark state , a $\bar{\Lambda}\Lambda$ ($^3S_1$) bound state \cite{phi_2170_exp11,phi_2170_exp12,phi_2170_exp13}, and a $\phi K\bar{K}$ resonance state \cite{phi_2170_exp14,phi_2170_exp15}. Given the diversity of these explanations, further theoretical and experimental studies are crucial to clarify the true nature of $\phi(2170)$.

In this work, we use the parton and hadron cascade model PACIAE 4.0 \cite{paciae_4} to simulate the production of $\phi(2170)$ in $e^+e^-$ collisions at $\sqrt{s}=4.95$ GeV. This energy is selected because BESIII has accumulated high-luminosity data at this point, where clear $\phi(2170)$ signals appear in the $\phi\eta$ and $\phi\eta'$ invariant mass spectra \cite{x_2300_BESIII}. In the PACIAE 4.0 model, the final partonic state (FPS) and the final hadronic state (FHS) are simulated and recorded sequentially. The $U(1)$ anomaly coupling allows non-strange quarks to couple to a vector $s\bar{s}$ component via soft-gluon interactions. This motivates us to explore the $d\bar{d}s\bar{s}$ tetraquark configuration in addition to the previously proposed $ss\bar{s}\bar{s}$ and $u\bar{u}s\bar{s}$ scenarios. We also investigate four other possibilities: an excited $s\bar{s}$ state, an $s\bar{s}g$ hybrid state, a $\bar{\Lambda}\Lambda$ bound state, and a $\phi K^+K^-$ resonance state.
The excited strangeonium, hybrid, and tetraquark candidates are formed by coalescing the constituent partons in the FPS using the dynamically constrained phase‑space coalescence (DCPC) model \cite{DCPC}, which is inspired by quantum statistical mechanics. Specifically, an $s\bar{s}$ pair coalesces into the strangeonium candidate, an $s\bar{s}g$ system into the hybrid candidate, and a four‑quark system ($u\bar{u}s\bar{s}$, $d\bar{d}s\bar{s}$ or $ss\bar{s}\bar{s}$) into the tetraquark candidate. By contrast, the $\bar{\Lambda}\Lambda$ bound state and the $\phi K^+K^-$ resonance state are produced via recombination of their constituent baryons ($\Lambda$ and $\bar{\Lambda}$) and mesons ($\phi$, $K^+$, $K^-$) in the FHS, respectively. For each $\phi(2170)$ candidate, we compute its orbital angular momentum $L$ in the rest frame and classify its spectrum using the standard $n^{2S+1}L_J$ notation \cite{book1}, where $n$, $S$, $L$, and $J$ denote the radial excitation number, total spin, orbital angular momentum, and total angular momentum, respectively. The $S$, $P$, $D$, $\dots$ wave shapes correspond to $L = 0, 1, 2, \dots$, respectively.

Given that $\phi(2170)$ has $J^{PC} = 1^{--}$ \cite{phi_2170}, it can be identified as either a $D$-wave $s\bar{s}$ state, a $P$-wave $s\bar{s}g$ hybrid state, a $P$-wave $u\bar{u}s\bar{s}/d\bar{d}s\bar{s}/ss\bar{s}\bar{s}$ tetraquark state, an $S$-wave $\bar{\Lambda}\Lambda$ bound state, or an $S$-wave $\phi K^+K^-$ resonance state. Although $X(2300)$ is regarded as a candidate for a $\phi\eta'/\phi\eta$ hadro-strangeonium state \cite{phi_2170_exp16}, $\phi(2170)$ cannot be regarded as such a state. The reason lies in the parity ($P$) and charge-conjugation parity ($C$) of a $\phi\eta'/\phi\eta$ system, which consists of a vector meson ($\phi$) and a pseudoscalar meson ($\eta$ or $\eta'$). For such a system, $P = (-1)^L$ and $C = (-1)^{L+1}$, where $L$ is the relative orbital angular momentum. Hence $P$ and $C$ always have opposite signs regardless of $L$. With $X(2300)$ having $1^{+-}$ and $\phi(2170)$ having $1^{--}$, only $X(2300)$ satisfies the $P$-$C$ relation of the $\phi\eta'/\phi\eta$ system. Consequently, $\phi(2170)$ cannot be a $\phi\eta'$ or $\phi\eta$ hadro-strangeonium state. In this work, we will compare the production yields, rapidity distributions, and transverse momentum spectra of the candidate configurations described above. The observed differences are expected to serve as key criteria for identifying the true nature of $\phi(2170)$.

\section{The model and methodology} \label{sec:method}

In the PACIAE 4.0 model, the simulation of $e^+e^-$ annihilation proceeds as follows. First, PYTHIA6 \cite{pythia_6} is executed with hadronization disabled and with the subsequent breaking of strings and diquarks/antidiquarks into quarks and antiquarks. To retain gluons for hybrid state formation, the splitting of gluons into quark-antiquark pairs is disabled. This yields an initial partonic state. Next, an energy deexcitation process is applied to energetic quarks (or antiquarks). The final partonic state (FPS) is then obtained after partonic $2\rightarrow 2$ rescattering, using leading-order perturbative QCD cross sections \cite{cs_1,cs_2}. The FPS contains many quarks, antiquarks, and gluons, each with four-dimensional coordinates and momenta. An intermediate hadronic state is produced by hadronizing the FPS via the Lund string fragmentation scheme \cite{pythia_6}. Finally, this intermediate state undergoes $2\rightarrow 2$ hadronic rescattering \cite{book2, paciae_3}, leading to kinetic freeze-out and generating the final hadronic state (FHS), which consists of abundant hadrons with their four-dimensional coordinates and momenta.

The DCPC model was originally proposed by us to study light nuclei production in proton-proton collisions at LHC energies \cite{DCPC}. Based on the final partonic or hadronic states generated by PACIAE, the DCPC model has been successfully applied to calculate the production yields of various exotic hadronic states, including $X(3872)$ \cite{ge2021, tai2023, she2024}, $Z_c^{\pm}(3900)$ \cite{zc_3900}, $G(3900)$ \cite{G_3900}, $P_c(4312)$, $P_c(4440)$, $P_c(4457)$ \cite{hui2022}, $\Omega_c^0$ \cite{Omega_c}, $T_{cs0}^{*}(2870)^{0}$ \cite{X_2870}, $X(2370)$ \cite{zhang2024}, and $X(2300)$ \cite{phi_2170_exp16}.

In this model, inspired by quantum statistical mechanics \cite{kobo1965, stowe2007}, the yield of a cluster consisting of $N$ particles is estimated by integrating over the phase space under a total energy constraint:
\begin{equation}
Y_N = \int\cdots\int_{E_{\alpha} \le E_{\text{tot}} \le E_{\beta}} \frac{d\boldsymbol{x}_1 d\boldsymbol{p}_1 \cdots d\boldsymbol{x}_N d\boldsymbol{p}_N}{h^{3N}},
\label{eq:yield_general}
\end{equation}
where $E_{\alpha}$ and $E_{\beta}$ are the lower and upper energy thresholds of the cluster, respectively, and $E_{\text{tot}} = \sum_{i=1}^N \sqrt{|\boldsymbol{p}_i|^2 + m_i^2}$ is the total energy of the cluster in the center-of-mass system (cms) of the $e^+e^-$ collision. The vectors $\boldsymbol{x}_i$ and $\boldsymbol{p}_i$ are the three-dimensional coordinate and momentum of the $i$-th constituent particle in the same cms.

A naturally formed cluster is assumed to satisfy specific dynamical constraints concerning the identities, coordinates, and momenta of its constituents. As an example, the yield of a $\phi(2170)$ resonance state composed of $\phi$, $K^+$, and $K^-$ is expressed as
\begin{equation}
Y_{\phi K^+K^-} = \int \delta_{123} \; \frac{d\boldsymbol{x}_1 d\boldsymbol{p}_1 \, d\boldsymbol{x}_2 d\boldsymbol{p}_2 \, d\boldsymbol{x}_3 d\boldsymbol{p}_3}{h^{9}},
\label{eq:yield_phikk}
\end{equation}
with the dynamical constraint
\begin{equation}
\delta_{123} = 
\begin{cases}
1, & \textrm {if}\ 1\equiv \phi,\ 1\equiv K^+,\ 3\equiv K^-,\\& R_i \le R_0\ (i=1,2,3), \\[2pt]
   &  m_0 - \Delta m \le m_{\text{inv}} \le m_0 + \Delta m, \\[6pt]
0, & \text{otherwise},
\end{cases}
\label{eq:delta_phikk}
\end{equation}
where $m_0 = 2164\ \text{MeV}/c^2$ is the nominal mass of the $\phi(2170)$ candidate \cite{PDG}, $\Delta m$ is the mass uncertainty (a free parameter taken as the decay width of $\phi(2170)$), and $R_0$ is the cluster radius (also a free parameter). Here $R_i = |\boldsymbol{x}_i^*|$, with $\boldsymbol{x}_i^*$ being the position vector of the $i$-th constituent meson ($\phi$, $K^+$, or $K^-$) in the rest frame of the cluster. To obtain $\boldsymbol{x}_i^*$, we Lorentz-transform the cms coordinate $\boldsymbol{x}_i$ to the cluster rest frame and then propagate each component meson freely from its own freeze-out time to the latest freeze-out time among the components \cite{zhull1,zhull2}. In the $\phi K^+K^-$ resonance interpretation, the $\phi(2170)$ decays into $\phi f_0(980)$, with $f_0(980)$ subsequently decaying into $K^+K^-$. Consequently, the cluster radius $R_0$ is naturally of the order of the sum of the radii of the $\phi$ meson and the $f_0(980)$ system. In this work, we take $1\ \text{fm} < R_0 < 2\ \text{fm}$. The invariant mass $m_{\text{inv}}$ is computed from the four‑momenta of the three constituent mesons:
\begin{equation}
m_{\text{inv}} = \sqrt{\left( \sum_{i=1}^{3} E_i \right)^2 - \left( \sum_{i=1}^{3} \boldsymbol{p}_i \right)^2},
\label{eq:inv_mass}
\end{equation}
(e.g., excited strangeonium, tetraquark, $s\bar{s}g$ hybrid, or $\bar{\Lambda}\Lambda$ bound states)where $E_i$ and $\boldsymbol{p}_i$ ($i=1,2,3$) are the energy and three‑momentum of the constituent meson ($\phi$, $K^+$, or $K^-$) in the $e^+e^-$ cms. The yields of $\phi(2170)$ candidates with other configurations (e.g., excited strangeonium, tetraquark, $s\bar{s}g$ hybrid, or $\bar{\Lambda}\Lambda$ bound states) are evaluated analogously, using appropriate parameters. The values of $\Delta m$ and $R_0$ for each scenario are listed in Table~\ref{tab:phi2170_para}.
 
\begin{table}[h]
\caption{The parameters of the mass uncertainty and the radius for the $\phi(2170)$ candidates of the excited $s\bar{s}$ strangeonium state, the $u\bar{u}s\bar{s}/d\bar{d}s\bar{s}/ss\bar{s}\bar{s}$ tetraquark state, the $s\bar{s}g$ hybrid state, the $\bar{\Lambda}\Lambda$ bound state, and the $\phi K^+K^-$ resonance state.}\label{tab:phi2170_para} 
\begin{ruledtabular}
\begin{tabular}{cccccc}
                  &     excited    $s\bar{s}$      & $s\bar{s}g$  & $u\bar{u}s\bar{s}/d\bar{d}s\bar{s}/ss\bar{s}\bar{s}$   & $\bar{\Lambda}\Lambda$         & $\phi K^+K^-$        \\
 \colrule                 
 $\Delta m$               &  \multirow{2}{*}{106} & \multirow{2}{*}{106}                   & \multirow{2}{*}{106}                  & \multirow{2}{*}{106}          & \multirow{2}{*}{106}          \\
 (MeV$/c^2$)                 &                    &                   &          &           \\
 \hline
$R_0$ (fm)                & 1.0            &  1.0    & 1.0                  & 1.0-2.0         & 1.0-2.0        
\end{tabular}
\end{ruledtabular}
\end{table}

To generate the $\phi(2170)$ $\phi K^+K^-$ resonance state as an example, the following procedure is adopted. First, a list of component mesons, including $\phi$, $K^+$, and $K^-$, is constructed using the FHS simulated by the PACIAE model. A triple-loop iteration is then performed over all mesons in this list. Each combination of one $\phi$, one $K^+$, and one $K^-$ that satisfies the constraints given in Eq.~(\ref{eq:delta_phikk}) is accepted as a $\phi(2170)$ candidate. The constituent mesons of each accepted candidate are subsequently removed from the list. The triple-loop process is repeated on the updated list until no mesons remain or no further valid candidates can be formed. The production of $\phi(2170)$ candidates with other configurations  is performed in a similar manner, using appropriate particle lists and constraints. The couplings for both the production and decay of $\phi(2170)$ are not explicitly provided, as they are effectively embedded through the geometric constraints, kinematic constraints, and coalescence mechanisms within the DCPC model.

To determine the orbital angular momentum quantum number of a $\phi(2170)$ candidate for spectral classification, we follow a method detailed in our previous studies \cite{G_3900,phi_2170_exp16}. A brief outline is given here. In the rest frame of the candidate, the orbital angular momentum (OAM) $\boldsymbol{l}^{*}$ is defined as the vector sum of the individual OAMs of its constituents:
\begin{equation}
\boldsymbol{l}^{*} = \sum_{i=1}^{N} \boldsymbol{x}_i^{*} \times \boldsymbol{p}_i^{*},
\label{eq:OAM_phi2170}
\end{equation}
where $N$ denotes the number of constituents in the specific configuration (e.g., $N=2$ for the $s\bar{s}$ or $\bar{\Lambda}\Lambda$ state,  $N=3$ for the $s\bar{s}g$ or $\phi K^+K^-$ state, and $N=4$ for the tetraquark state), and $\boldsymbol{p}_i^{*}$ is the three‑momentum of the $i$-th constituent in the candidate's rest frame, obtained by Lorentz‑transforming the corresponding momentum $\boldsymbol{p}_i$ from the $e^+e^-$ cms. According to quantum mechanics, the squared orbital angular momentum is quantized:
\begin{equation}
|\boldsymbol{l}^{*}|^{2} = L(L+1)\hbar^{2},
\label{eq:OAM_quantum}
\end{equation}
with $\hbar$ the reduced Planck constant and $L$ the orbital angular momentum quantum number of the $\phi(2170)$ candidate. Since $L$ must be an integer, Eq.~(\ref{eq:OAM_quantum}) leads to:
\begin{equation}
L = \operatorname{round}\left( \frac{-1 + \sqrt{1 + 4|\boldsymbol{l}^{*}|^{2}/\hbar^{2}}}{2} \right),
\label{eq:OAM_extract}
\end{equation}
where $\operatorname{round}(X)$ returns the integer nearest to $X$.

Different theoretical interpretations of $\phi(2170)$ lead to distinct $P$ and $C$ formulas. If $\phi(2170)$ is interpreted as an excited $s\bar{s}$ state, the parity is $P = P_s \cdot P_{\bar{s}} \cdot (-1)^L = (-1)^{L+1}$, where $L$ is the total orbital angular momentum of the system.  For the tetraquark scenario, the intrinsic parity is $P = P_{u/d/s} \cdot P_{\bar{u}/\bar{d}/\bar{s}}\cdot  P_s  \cdot P_{\bar{s}} \cdot (-1)^L = (-1)^L$. In the $\bar{\Lambda}\Lambda$ bound state interpretation, the parity is $P = P_{\Lambda}\cdot P_{\bar{\Lambda}}\cdot(-1)^{L}=(-1)^{L+1}$. For the  strangeonium, tetraquark, and $\bar{\Lambda}\Lambda$ configurations, the $C$-parity follows the common form: $C = (-1)^{L + S}$, where $S$ denotes the total spin of the constituent particles. In the $s\bar{s}g$ hybrid interpretation, the parity is given by $P = P_s \cdot P_{\bar{s}} \cdot P_g  \cdot (-1)^{L_{s\bar{s}}+L_g}=(-1)^{L_{s\bar{s}}+L_g}=(-1)^L$, where $L = L_{s\bar{s}} + L_g$, $L_g$ is the relative orbital angular momentum between the gluon and the $s\bar{s}$ center of mass, and $L_{s\bar{s}}$ is the relative orbital angular momentum between the strange quark and the anti-strange quark \cite{phi_2170_exp3}. The $C$-parity for such a system becomes $C =C_{s\bar s}\cdot C_g=(-1)^{L_{s\bar{s}}+S_{s\bar{s}}+1}$, where  $S_{s\bar{s}}$ is the total spin of $s\bar{s}$ \cite{phi_2170_exp3, hybrid_C_parity}. Finally, in the $\phi K^+K^-$ resonance picture, the parity is $P = P_\phi \cdot P_{K^+} \cdot P_{K^-} \cdot (-1)^{L_{K^+K^-} + L_{\phi}} = (-1)^{L+1}$, where $P_{\phi}=-1$, $P_{K^\pm}=-1$, $L = L_{K^+K^-} + L_{\phi}$, $L_{K^+K^-}$ is the relative orbital angular momentum between $K^+$ and $K^-$, and $L_{\phi}$ is the relative orbital angular momentum  between $\phi$  and the $K^+K^-$ center of mass. The $C$-parity is $C =C_\phi \cdot C_{K^+K^-}=(-1)^{L_{K^+K^-}+S_{K^+K^-}+1}$, where $S_{K^+K^-}=0$ is the total spin of $K^+K^-$.
For the strangeonium, tetraquark, or $\bar{\Lambda}\Lambda$ configuration, with a given orbital angular momentum $L$ and total spin $S$ of the constituents, the total angular momentum $J$ of the $\phi(2170)$ candidate can take the values
\begin{equation}
J = |L - S|,\ |L - S| + 1,\ \dots,\ L + S.
\end{equation}
For the $s\bar{s}g$ hybrid state, since the gluon carries an intrinsic spin $S_g=1$, we first couple the total angular momentum of $s\bar{s}$, $\boldsymbol{J}_{s\bar{s}} = \boldsymbol{L}_{s\bar{s}} + \boldsymbol{S}_{s\bar{s}}$, with the gluon spin to form an intermediate momentum $\boldsymbol{J}_{\text{int}} = \boldsymbol{J}_{s\bar{s}} + \boldsymbol{1}$, where $J_{s\bar{s}}$ runs over all integer values from $|L_{s\bar{s}}-S_{s\bar{s}}|$ to $L_{s\bar{s}}+S_{s\bar{s}}$. The final total angular momentum is then obtained by coupling $\boldsymbol{J}_{\text{int}}$ with the gluon's relative orbital angular momentum $\boldsymbol{L}_g$:
\begin{equation}
J = |J_{\text{int}} - L_g|,\ |J_{\text{int}} - L_g| + 1,\ \dots,\ J_{\text{int}} + L_g,
\end{equation}
where $J_{\text{int}}$ takes all integer values from $|J_{s\bar{s}} - 1|$ to $J_{s\bar{s}} + 1$. The total angular momentum for the $\phi K^+K^-$ resonance state is determined in a similar way.
The experimentally determined quantum numbers of $\phi(2170)$ are $J^{PC}=1^{--}$, which imposes constraints on the allowed $(L,S)$ combinations for each interpretation.

\begin{table}[]
\caption{The $J^{PC}$s for the $S$-, $P$-, and $D$-wave $\phi(2170)$ candidates with different configurations.}\label{tab:phi2170_JPC}
\begin{ruledtabular}
\begin{tabular}{ccccc}
                  &     & $S$-wave & $P$-wave & $D$-wave \\ \hline
\multirow{2}{*}{$s\bar s$} & $S=0$ & $0^{-+}$  & $1^{+-}$  & $2^{-+}$   \\
                  & $S=1$ &  $1^{--}$ & $(0,1,2)^{++}$  &  $(1,2,3)^{--}$ \\  \hline
\multirow{2}{*}{$s\bar{s}g$} & $S_{s\bar{s}}=0$ &  $1^{+-}$ & $(0,1,2)^{-\pm}$  &  $(0,1,2,3)^{+\pm}$ \\
                  & $S_{s\bar{s}}=1$ & $(0,1,2)^{++}$  & $(0,1,2,3)^{-\pm}$  &  $(0,1,2,3,4)^{+\pm}$ \\  \hline
$u\bar{u}s\bar{s}$, & $S=0$ &  $0^{++}$ & $1^{--}$  &  $2^{++}$ \\
$d\bar{d}s\bar{s},$          & $S=1$ & $1^{+-}$  & $(0,1,2)^{-+}$  &  $(1,2,3)^{+-}$ \\
or $ss\bar{s}\bar{s}$        & $S=2$ & $2^{++}$  & $(1,2,3)^{--}$  &  $(0,1,2,3,4)^{++}$ \\  \hline
\multirow{2}{*}{$\bar{\Lambda}\Lambda$} & $S=0$ & $0^{-+}$  & $1^{+-}$  & $2^{-+}$   \\
                  & $S=1$ &  $1^{--}$ & $(0,1,2)^{++}$  &  $(1,2,3)^{--}$ \\  \hline
$\phi K^+K^-$  & $S_{K^+K^-}=0$ & $1^{--}$  & $(0,1,2)^{+\pm}$  & $(0,1,2,3)^{-\pm}$ 
\end{tabular}
\end{ruledtabular}
\end{table}

\begin{table}[]
\caption{The event-average yields of the $D$-wave  $s\bar s$ strangeonium state, the $P$-wave $s\bar sg$ hybrid state, the $P$-wave $u\bar{u}s\bar{s}/d\bar{d}s\bar{s}/ss\bar s\bar s$ tetraquark state, the $S$-wave $\bar{\Lambda}\Lambda$ bound state, and the $S$-wave $\phi K^+K^-$ resonance  state   for the $\phi(2170)$ candidates in $e^+e^-$ collisions at $\sqrt{s} = 4.95\text{ GeV}$. The uncertainties quoted are statistical errors. }\label{tab:phi_2170_yield}
\begin{ruledtabular}
\begin{tabular}{ccc}
      & $D-$wave $s\bar s$  & $P-$wave $s\bar sg$\\\hline
yield &  $(1.223\pm0.002)\times 10^{-4}$ & $(1.718\pm0.003)\times 10^{-4}$  \\\hline
      &  $P-$wave $u\bar{u}s\bar{s}$  &$P-$wave $d\overline{d}s\bar{s}$\\\hline
yield & $(1.595\pm0.003)\times 10^{-4}$  & $(1.313\pm0.002)\times 10^{-4}$   \\\hline
      &$P-$wave $ss\bar  s\bar s$ &$S-$wave $\overline{\Lambda}\Lambda$  \\\hline
yield &  $(7.888\pm0.060)\times 10^{-6}$& $(6.575\pm0.017)\times 10^{-5}$   \\\hline
      & $S-$wave $\phi K^+K^-$ &  \\\hline
yield & $(6.426\pm0.017)\times 10^{-5}$ &  
\end{tabular}
\end{ruledtabular}
\end{table}

\section{Results and discussions} 
The $e^+e^-$ collisions at $\sqrt{s} = 4.95\text{ GeV}$ are simulated using the PACIAE 4.0 model \cite{paciae_4}. A total of 2.2 billion events are generated with the default model parameters. The excited strangeonium, hybrid, and tetraquark states of the $\phi(2170)$ candidates are, respectively, produced by coalescing the partons $s\bar{s}$, $s\bar{s}g$, and $u\bar{u}s\bar{s}/d\bar{d}s\bar{s}/ss\bar{s}\bar{s}$ in the FPS with the DCPC model.
The $\bar{\Lambda}\Lambda$ bound state and the $\phi K^+K^-$ resonance state are generated via recombination of their constituent baryons ($\Lambda$ and $\bar{\Lambda}$) and mesons ($\phi$, $K^+$, and $K^-$) in the FHS with the DCPC model.  Table~\ref{tab:phi2170_JPC} lists the $J^{PC}$ quantum numbers for the $S$-, $P$-, and $D$-wave $\phi(2170)$ candidates under various configurations. They are determined following the procedure described in Section~\ref{sec:method}. Given that $\phi(2170)$ has quantum numbers $J^{PC}=1^{--}$ \cite{phi_2170}, several interpretations are possible: a $D$-wave $s\bar{s}$ state, a $P$-wave $s\bar{s}g$ hybrid, a $P$-wave $u\bar{u}s\bar{s}/d\bar{d}s\bar{s}/ss\bar{s}\bar{s}$ tetraquark, an $S$-wave $\bar{\Lambda}\Lambda$ bound state, or an $S$-wave $\phi K^+K^-$ resonance state. For the excited strangeonium interpretation, an $S$-wave $3^3S_1$ configuration has also been proposed \cite{phi_2170_exp2, phi_2170_exp1_1,phi_2170_exp1_2}. However, as pointed out in Refs.~\cite{phi_2170_2, phi_2170_exp1}, this $3^3S_1$ assignment is disfavored because it predicts a rather large width of about 380 MeV \cite{phi_2170_exp1_1}, which exceeds the experimental upper limit of about 100 MeV \cite{phi_2170_1,phi_2170_8}. In contrast, the $2^3D_1$ assignment yields a total width of approximately 186 MeV, which is consistent with the estimate in Ref.~\cite{phi_2170_2}. Therefore, for the excited strangeonium scenario, we retain only the $D$-wave configuration. For the $\bar{\Lambda}\Lambda$ bound state and the $\phi K^+K^-$ resonance, $D$-wave configurations could also yield $J^{PC}=1^{--}$. Nevertheless, as discussed in Refs.~\cite{phi_2170_exp11,phi_2170_exp12,phi_2170_exp13,phi_2170_exp14,phi_2170_exp15}, $\phi(2170)$ is most favorably interpreted as a $^3S_1$ $\bar{\Lambda}\Lambda$ bound state or an $S$-wave $\phi K^+K^-$ resonance. Consequently, $D$-wave configurations for these two scenarios are not considered in this work.

The event-average yields of the $D$-wave $s\bar s$ strangeonium state, the $P$-wave $s\bar sg$ hybrid state, the $P$-wave $u\bar{u}s\bar{s}/d\bar{d}s\bar{s}/ss\bar s\bar s$ tetraquark state, the $S$-wave $\bar{\Lambda}\Lambda$ bound state, and the $S$-wave $\phi K^+K^-$ resonance  state   for the $\phi(2170)$ candidates are estimated for the first time and summarized in  Table \ref{tab:phi_2170_yield}.  We observe that the yields of the $D$-wave $s\bar{s}$, the $P$-wave $s\bar{s}g$, as well as the $P$-wave $u\bar{u}s\bar{s}$ and $d\bar{d}s\bar{s}$ states are of the order of $10^{-4}$; the yields for the $S$-wave $\bar{\Lambda}\Lambda$ and $\phi K^+K^-$ states are of the order of $10^{-5}$; while the yield for the $P$-wave $ss\bar{s}\bar{s}$ state is of the order of $10^{-6}$. Moreover, the yield of the $P$-wave $u\bar{u}s\bar{s}$ state is slightly larger than that of the $d\bar{d}s\bar{s}$ state. The higher yield of $u\bar{u}s\bar{s}$ compared to $d\bar{d}s\bar{s}$ originates from the higher charge production probability and lower production threshold of the $u$ quark in the initial hard scattering process. This initial advantage is suppressed during parton rescattering and the dynamically constrained phase-space coalescence, ultimately yielding a ratio of about 1.23:1. Furthermore, the $ss\bar{s}\bar{s}$ yield is about two orders of magnitude lower than that of $u\bar{u}s\bar{s}$, which is a combined result of the suppression of strange quark initial production and the difficulty of recombining multiple rare quark pairs.

\begin{figure}[]
\centering
\includegraphics[scale=0.461]{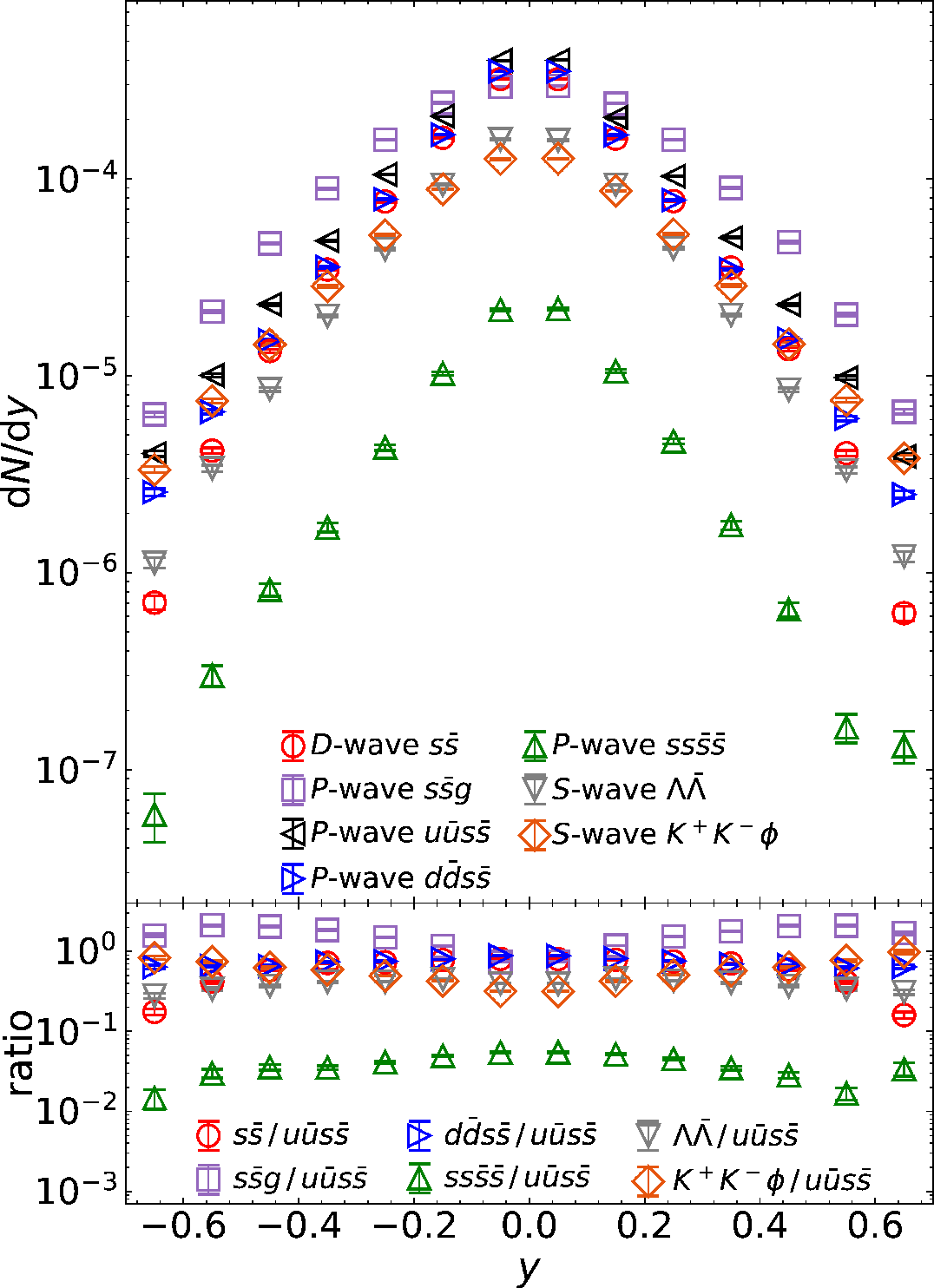}
\caption{\label{fig:yy_phi2170}  Upper panel: the simulated $y$ 
single-differential distributions for the $D$-wave $s\bar{s}$ (circles), the $P$-wave $s\bar{s}g$ (squares), $u\bar{u}s\bar{s}$ (left triangles),  $d\bar{d}s\bar{s}$ (right triangles), and $ss\bar{s}\bar{s}$ (upward triangles), as well as the  $S$-wave $\bar{\Lambda}\Lambda$ (downward triangles) and $\phi K^+K^-$ (diamonds) states of the $\phi(2170)$ candidates in $e^+e^-$ collisions at $\sqrt{s}$= 4.95 GeV. Lower panel: the ratios between two distributions denoted by legend. The error bars represent the statistical uncertainties.}
\end{figure}

\begin{figure}
\centering
\includegraphics[scale=0.461]{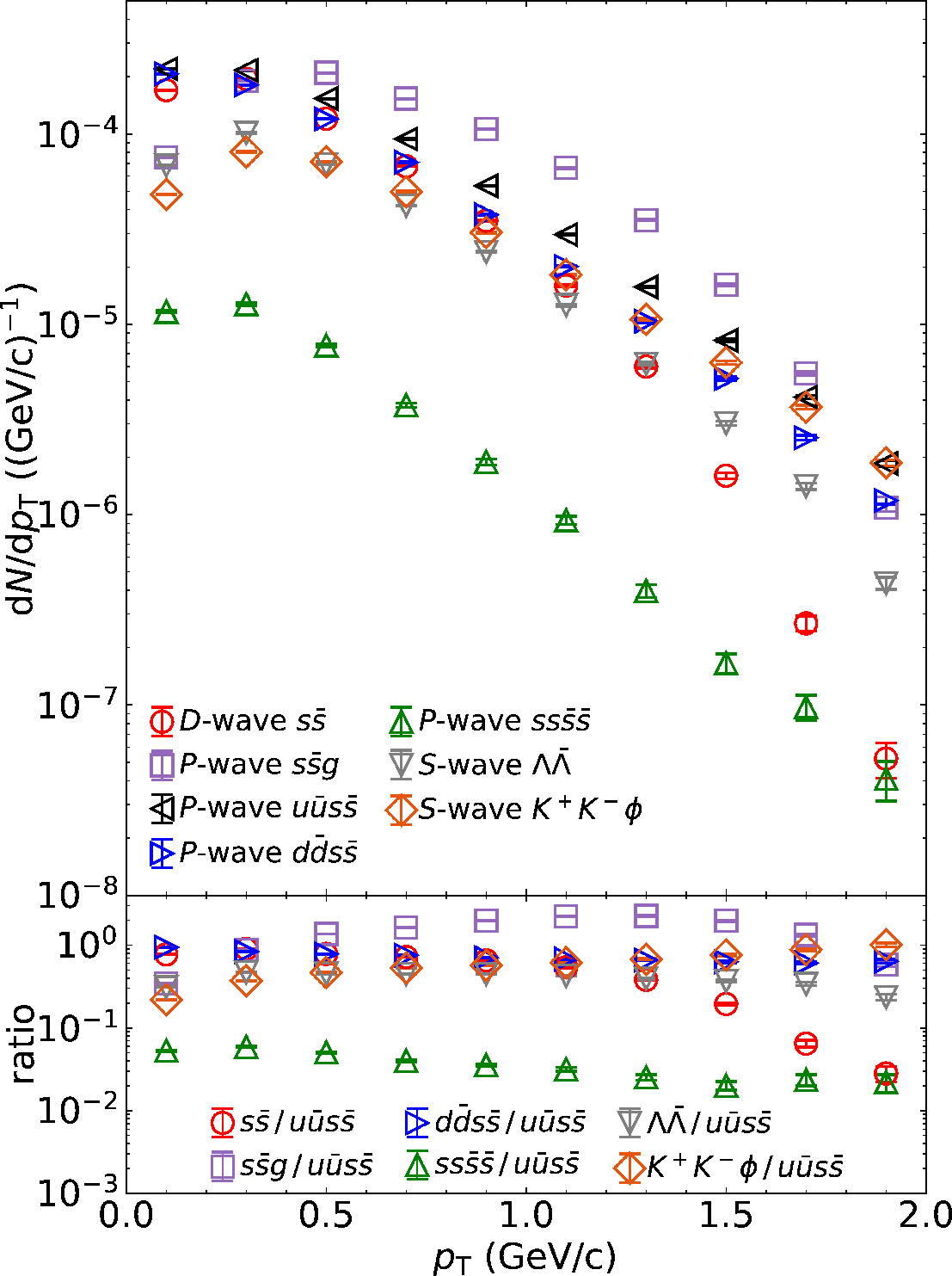}
\caption{\label{fig:pt_phi2170} Similar as that in Fig. \ref{fig:yy_phi2170}, but for the $p_{\rm T}$-differential cross sections of  the $\phi(2170)$ candidates with different configurations in $e^+e^-$ collisions at $\sqrt{s}$= 4.95 GeV.}
\end{figure}

We have investigated the influence of the coalescence radius $R_0$ on the production yields of the $\phi(2170)$ candidates in various configurations. For the strangeonium and tetraquark states, the yields exhibit a strong dependence on the upper bound of $R_0$. Specifically, when the upper limit is reduced from 1.0 fm to 0.7 fm, the yields of the $D$-wave $s\bar{s}$ and $P$-wave $s\bar{s}g$ states drop from $1.22\times 10^{-4}$ and $1.72\times 10^{-4}$ to $4.44\times 10^{-6}$ and $4.51\times 10^{-6}$, respectively. Similarly, the yields of the $P$-wave $u\bar{u}s\bar{s}$, $d\bar{d}s\bar{s}$, and $ss\bar{s}\bar{s}$ states decrease from $1.60\times 10^{-4}$, $1.31\times 10^{-4}$, and $7.89\times 10^{-6}$ to $1.61\times 10^{-6}$, $1.21\times 10^{-6}$, and $3.91\times 10^{-8}$, respectively. In contrast, the $S$-wave $\bar{\Lambda}\Lambda$ and $\phi K^+K^-$ states show an opposite tendency with respect to the lower bound of $R_0$. When the lower limit is decreased from 1.0 fm to 0.7 fm, their yields increase from $6.58\times 10^{-5}$ and $6.43\times 10^{-5}$ to $1.30\times 10^{-4}$ and $1.46\times 10^{-4}$, respectively. Moreover, further increasing the upper bound from 2.0 fm to 2.3 fm raises their yields to $8.63\times 10^{-5}$ and $8.18\times 10^{-5}$, respectively. We have also studied the dependence of the yields on the mass uncertainty $\Delta m$. When $\Delta m$ is reduced from the full decay width of $\phi(2170)$ ($106$ MeV/$c^2$) to half of it ($53$ MeV/$c^2$), the yields of the $D$-wave $s\bar{s}$, the $P$-wave $s\bar{s}g$, $u\bar{u}s\bar{s}$, $d\bar{d}s\bar{s}$, $ss\bar{s}\bar{s}$, and the $S$-wave $\phi K^+K^-$ states decrease by factors of approximately 2.03, 2.00, 1.89, 1.90, 1.89, and 1.90, respectively. For the $S$-wave $\bar{\Lambda}\Lambda$ state, the yield drops to zero. We have verified that the minimum value of $\Delta m$ allowing the existence of $\bar{\Lambda}\Lambda$ candidates is $0.64$ times the total width of $\phi(2170)$.
Furthermore, we examined the sensitivity of the results to the method used for extracting the orbital angular momentum quantum number $L$ in Eq.~(\ref{eq:OAM_extract}). An alternative procedure based on “trunc($X$)”, which discards the fractional part of $X$, was compared with the default “round($X$)” method. Replacing “round” with “trunc” enhances the yields of the  $P$-wave $s\bar{s}g$, $S$-wave $\bar{\Lambda}\Lambda$, and $S$-wave $\phi K^+K^-$ states by factors of roughly 1.23, 1.45, and 2.20, respectively. By contrast, the yields of the $D$-wave $s\bar{s}$ as well as  the $P$-wave $u\bar{u}s\bar{s}$, $d\bar{d}s\bar{s}$, and $ss\bar{s}\bar{s}$ tetraquark states are reduced by factors of about 1.07, 1.11, 1.13, and 1.71, respectively.

The upper panels in Figs. \ref{fig:yy_phi2170} and  \ref{fig:pt_phi2170} show, respectively, the simulated $y$ and $p_{\rm T}$ single-differential distributions for the $D$-wave $s\bar{s}$ (circles), the $P$-wave $s\bar{s}g$ (squares), $u\bar{u}s\bar{s}$ (left triangles),  $d\bar{d}s\bar{s}$ (right triangles), and $ss\bar{s}\bar{s}$ (upward triangles), as well as the  $S$-wave $\bar{\Lambda}\Lambda$ (downward triangles) and $\phi K^+K^-$ (diamonds) states of the $\phi(2170)$ candidates in $e^+e^-$ collisions at $\sqrt{s}$= 4.95 GeV. The lower panels of both figures show the corresponding ratios of the distributions for the $D$-wave $s\bar{s}$, the $P$-wave $s\bar{s}g$, $d\bar{d}s\bar{s}$, and $ss\bar{s}\bar{s}$, as well as the $S$-wave $\bar{\Lambda}\Lambda$ and $\phi K^+K^-$ states to that of the $P$-wave $u\bar{u}s\bar{s}$ state. In the upper panels of Fig. \ref{fig:yy_phi2170}, there is a peak at mid-rapidity for the rapidity distributions of the $\phi(2170)$ candidates with different configurations. This can be understood as follows: the central rapidity region collects particles from all event types regardless of jet direction, while the large-rapidity region receives contributions only from events with jets aligned close to the beam axis. Consequently, the yield at central rapidity is significantly enhanced, producing a well-defined peak.  Furthermore, the peak height follows the hierarchy $u\bar{u}s\bar{s}> d\bar{d}s\bar{s}>s\bar{s}>s\bar{s}g> \bar{\Lambda}\Lambda> \phi K^+K^->ss\bar{s}\bar{s} $. Although the $s\bar{s}g$ configuration has the largest yield, its peak height is not the highest among all configurations. This is because the rapidity distribution of the $s\bar{s}g$ state is broader than those of the other configurations, which lowers its peak height. In the upper panel of Fig. \ref{fig:pt_phi2170}, there is a peak at low $p_{\rm T}$ for the $p_{\rm T}$ single-differential distributions of the $\phi(2170)$ candidates with different configurations. While the $D$-wave $s\bar{s}$, the $P$-wave $u\bar{u}s\bar{s}$, $d\bar{d}s\bar{s}$, and $ss\bar{s}\bar{s}$, as well as the $S$-wave $\bar{\Lambda}\Lambda$ and $\phi K^+K^-$ states all exhibit a peak around $p_{\rm T} \simeq 0.3$ GeV/$c$, the $P$-wave $s\bar{s}g$ hybrid state shows a harder spectrum with its peak shifted to $p_{\rm T} \simeq 0.5$ GeV/$c$. This shift can be attributed to the different origins of transverse momentum. For the pure quark systems ($s\bar{s}$ and tetraquarks), the $p_{\rm T}$ is inherited from the relatively soft transverse momentum distribution of the constituent quarks produced in the parton shower, which typically peaks around $0.3$ GeV/$c$. In contrast, the $s\bar{s}g$ hybrid state contains a gluon, and 
gluons in parton cascades have intrinsically harder $p_{\rm T}$ spectra due to their stronger radiation and self-coupling \cite{gluon_broadening}. This leads to a broader $p_{\rm T}$ distribution for the hybrid state and shifts its peak to a higher value of $0.5$ GeV/$c$. For the $S$-wave $\bar{\Lambda}\Lambda$ and $\phi K^+K^-$ states, which are formed via recombination in the FHS, the transverse momentum spectra reflect the intrinsic $p_{\rm T}$ distributions of the constituent hadrons produced by the Lund string fragmentation.  In particular, the $\bar{\Lambda}\Lambda$ state involves heavy baryons. Due to their larger masses, these baryons are produced with intrinsically softer transverse momentum spectra in the fragmentation process, leading to a lower effective temperature and thus a lower peak position. The $\phi K^+K^-$ state, on the other hand, is a three-body system. Its formation via recombination in the FHS requires the three constituent momenta to sum to a total invariant mass within the narrow resonance window of $\phi(2170)$. High-$p_{\rm T}$ components tend to push the invariant mass outside this window and are therefore kinematically suppressed. As a result, both of these states naturally exhibit soft $p_{\rm T}$ spectra with peaks around $0.3$ GeV/$c$.

The observed differences in the yields, rapidity distributions, and $p_{\rm T}$ spectra among the various configurations of the $\phi(2170)$ candidates may provide valuable criteria for unraveling the nature of $\phi(2170)$. We suggest that these observables be measured experimentally in $e^+e^-$ collisions at BESIII energies and compared with our present predictions.

\section{acknowledgments}
This work is supported by the National Natural Science
Foundation of China under grant Nos. 11447024, 11505108 and 12375135, and by the 111 project of the
foreign expert bureau of China. Y.L.Y. acknowledges the financial support
from Key Laboratory of Quark and Lepton Physics in Central
China Normal University under grant No. QLPL201805 and the Continuous Basic
Scientific Research Project (No, WDJC-2019-13). W.C.Z. is supported
by the Natural Science Basic Research Plan in Shaanxi Province of China
(No. 2023-JCYB-012). H.Z. acknowledges the financial support from
Key Laboratory of Quark and Lepton Physics in Central China Normal University
under grant No. QLPL2024P01.


\end{document}